\documentstyle[pra,aps,epsf]{revtex}

\newcommand{\beq}{\begin{equation}}
\newcommand{\eeq}{\end{equation}}
\newcommand{\bea}{\begin{eqnarray}}
\newcommand{\eea}{\end{eqnarray}}
\newcommand{\beao}{\begin{eqnarray*}}
\newcommand{\eeao}{\end{eqnarray*}}

\begin{document}
\draft
\thispagestyle{empty}
\twocolumn[\hsize\textwidth\columnwidth\hsize\csname@twocolumnfalse%
\endcsname
\title{
Casimir force at both non-zero temperature and
finite conductivity
}
\author{
M.~Bordag,${}^1$
B.~Geyer,${}^2$
G.~L.~Klimchitskaya,${}^3$
 and V.~M.~Mostepanenko${}^4$}

\address
{Institute for Theoretical Physics, Leipzig
University,  Augustusplatz 10/11, 04109 Leipzig, 
Germany\\
${}^1${Electronic address: Michael.Bordag@itp.uni-leipzig.de}\\
${}^2${Electronic address: geyer@rz.uni-leipzig.de}\\
${}^3${on leave from North-West Polytechnical 
Institute,St.Petersburg, Russia.\\  
Electronic address:  galina@GK1372.spb.edu}\\
${}^4${on leave from A.Friedmann Laboratory
for Theoretical  Physics, St.Petersburg, Russia.\\ 
Electronic  address: mostep@fisica.ufpb.br
}}
\maketitle
\begin{abstract}
We find the joint effect of non-zero temperature and finite conductivity
onto the Casimir force between real metals. Configurations of two
parallel plates and a sphere (lens) above a plate are considered.
Perturbation theory in two parameters (the relative temperature and
the relative penetration depth of zero point oscillations into the metal)
is developed. Perturbative results are compared with computations.
Recent evidence concerning possible existence of large temperature
corrections at small separations between the real metals is not supported. 
\end{abstract}

\pacs{12.20.Ds, 11.10.Wx, 12.20.Fv}
]

Currently the Casimir effect is attracting considerable interest.
A large amount of information is available nowadays
of both theoretical and experimental nature. 
Theoretical progress was made in elaborating different approximate
methods [1--3] and in the problem of a dielectric sphere where, for
instance, the structure of the ultraviolet divergencies had been
clarified \cite{4}. In \cite{5} the additive method was successfully
applied to a dilute dielectric ball, and in \cite{6}  
a progress was made in obtaining analytical results. 
The Casimir force was demonstrated between metallic 
surfaces of a spherical lens (sphere) above a disk 
using a torsion pendulum \cite{10} and an atomic force microscope 
\cite{11,13}.

The increased accuracy of Casimir force measurements invites a further
investigation of different theoretical corrections. In \cite{14} the
Casimir force for the configuration of a sphere above a plate was
computed by taking into account surface roughness and finite conductivity
corrections up to fourth order in respective small parameters. That result
is in excellent agreement with the measured Casimir
force. Except for contributions of surface roughness and finite
conductivity, corrections due to non-zero temperature play a dominant
role above some distance between the test bodies. The general
expression for the temperature Casimir force between dielectric plates
was firstly obtained in \cite{15} (see also \cite{16}). The temperature
Casimir force between perfectly conducting plates was found in 
[13--15], including the limiting cases of large and small plate
separations (high and low temperatures). These results were modified
for the configuration of a spherical lens above a disk in \cite{10}.
The temperature corrections are found to be insignificant within the
separations of experiments \cite{11,13} (from $a\approx$0.1\,$\mu$m till
$a$=0.9\,$\mu$m or 0.5\,$\mu$m). As for experiment \cite{10} they
constitute up to 174\% of the net force at room temperature $T$=300\,K
at the largest separation $a$=6\,$\mu$m \cite{20} (in spite of this,
experimental data of \cite{10} are not sufficiently accurate to
demonstrate temperature corrections). In particular, it should be 
emphasized that the joint effect of non-zero temperature and finite
conductivity of the boundary metal was not investigated up to the present.

The computations of the recent paper \cite{21} have cast some doubt 
on the possibility to use the low temperature limit of the temperature
corrections for perfect conductors [13--15] in order to describe
the real metals. For the configuration of a sphere above a disk the
difference of 4\,pN at $a\sim$0.1\,$\mu$m was found depending on whether
one uses the expression for the temperature Casimir force for a real metal
or the zero temperature one. This is in contradiction with the generally
accepted behavior of the temperature correction between perfect conductors
at small separations which is proportional to 
$(k_B Ta/\hbar c)^3$ for a sphere above a disk,
$k_B$ being the Bolzmann constant [7,13--15].
The authors of \cite{21} hypothesized that for a real conductor the
temperature correction to the Casimir force at low temperature can
behave as $k_B Ta/\hbar c$ and be important.

Here we present a perturbative calculation          
of the joint influence of non-zero temperature and
finite conductivity on the Casimir force. The obtained
results are the generalization of [13--15] to the case of real metals. 
They can be used for
the interpretation of precision experiments on Casimir force. No
unexpected large temperature contributions arise at small separations.
The erroneous assumption of \cite{21} is explained below in detail.

We start with the configuration of two plane parallel plates with the
dielectric permittivity $\varepsilon$ separated by an empty gap of thickness
$a$. At arbitrary temperature $T$ the attractive force per unit area 
acting between plates is given by the Lifshitz formula \cite{16}
\beq
F_{pp}(a)=-\frac{k_B T}{\pi c^3}
\sum\limits_{n=0}^{\infty}{\vphantom{\sum}}^{\prime}\,
\xi_n^3
\int\limits_{1}^{\infty}p^2dp\left(Q_1^{-1}+Q_2^{-1}\right),
\label{1}
\eeq
\noindent
where
\bea
&&
Q_1=\frac{(s+p\varepsilon)^2}{(s-p\varepsilon)^2}
e^{\frac{2p\xi_n}{c}a}-1,\quad
Q_2=\frac{(s+p)^2}{(s-p)^2}
e^{\frac{2p\xi_n}{c}a}-1,
\nonumber\\
&&s=\sqrt{\varepsilon -1+p^2},\quad
\xi_n=\frac{2\pi k_B T}{\hbar}n, \quad
\varepsilon\equiv\varepsilon(i\xi_n).
\label{2}
\eea
\noindent
The prime on the sum indicates that the term with $n=0$ is 
to be taken with
the coefficient 1/2. Let us now introduce new variables
$x_n=2a\xi_n/c$ and $z=x_n p$. It is evident that
$x_n=\tau n\equiv 2\pi n T/T_{\mbox{\scriptsize{eff}}}$, 
where the effective
temperature is defined by 
$k_B T_{\mbox{\scriptsize{eff}}}=\hbar c/(2a)$ \cite{22}.
In new variables Eq.~(\ref{1}) takes the form
\bea
&&F_{pp}(a)=-\frac{k_B T}{8\pi a^3}
\sum\limits_{n=0}^{\infty}{\vphantom{\sum}}^{\prime}\,
\varphi_{pp}(x_n),
\label{3}\\
&&
\varphi_{pp}(x_n)=
\int\limits_{x_n}^{\infty}
z^2dz           \left(Q_1^{-1}+Q_2^{-1}\right),
\nonumber
\eea
\noindent
where $Q_{1,2}$ in (\ref{2}) are expressed now in terms of $x_n,\,z$.

The sum in (\ref{3}) can be calculated with the help of the Abel-Plana
formula \cite{22}. The result is
\bea
&&F_{pp}(a)=-\frac{k_B T}{8\pi a^3}
\left[\frac{1}{\tau}
\int\limits_{0}^{\infty}dx
\int\limits_{x}^{\infty}
z^2dz\left(Q_1^{-1}+Q_2^{-1}\right)
\right.\nonumber\\
&&\phantom{aaaa}+i\left.
\int\limits_{0}^{\infty}
\frac{\varphi_{pp}(i\tau y)-\varphi_{pp}(-i\tau y)}{e^{2\pi y}-1}dy\right].
\label{4}
\eea

The first term in the right-hand side of (\ref{4}) is the Casimir force
at zero temperature, the second one takes into account the temperature
corrections. The zero temperature contribution was calculated in \cite{23}
numerically by the use of optical tabulated data for the complex
refractive index (an alternative computation \cite{24} contains some errors
which are indicated in \cite{23}). Independently, in \cite{25} it was  
determined by perturbation theory up to the fourth order in the small 
parameter $\delta_0/a$ ($\delta_0$ being the effective penetration depth of
electromagnetic zero point oscillations into the metal). Thereby, the
plasma model was used for the dielectric permittivity 
\beq
\varepsilon(i\xi)=1+\frac{\omega_p^2}{\xi^2}=
1+\frac{\tilde\omega_p^2}{x^2},
\label{5}
\eeq
\noindent
where $\omega_p$ is the effective plasma frequency, and
$\tilde\omega_p=2a\omega_p/c$, so that 
$\alpha\equiv 1/\tilde\omega_p=\delta_0/(2a)$. The results of \cite{23}
and \cite{25} are in good agreement for space separations
$a\geq \lambda_p=2\pi c/\omega_p$.
It is well known that the plasma model does not take into account
the contribution of relaxation processes which are taken into
consideration by the Drude model (see below). However, the variation 
of the Casimir force obtained by both models remains smaller than
2\% \cite{23}.

Let us calculate the second term of (\ref{4}) in the application
range of plasma model and under the condition 
$T\ll T_{\mbox{\scriptsize{eff}}}$.
To do this we use the representation of (\ref{3})
\beq
\varphi_{pp}(x)=\varphi_{pp}^{(1)}(x)+\varphi_{pp}^{(2)}(x)
\equiv
\int\limits_{x}^{\infty}\!\!
z^2dzQ_1^{-1}+
\int\limits_{x}^{\infty}\!\!
z^2dzQ_2^{-1}.
\label{6}
\eeq
\noindent
Introducing $x_0<1$ one can write
\beq
\varphi_{pp}^{(i)}(x)=
\int\limits_{x}^{x_0}
z^2dzQ_i^{-1}+
\int\limits_{x_0}^{\infty}
z^2dzQ_i^{-1}.
\label{7}
\eeq
\noindent
Considering firstly the case $i=2$, we notice that in the plasma model 
(\ref{5})
the second term from the right-hand side of (\ref{7}) does not depend on
$x$. Expanding $Q_2^{-1}$ from the first term of (\ref{7}) into a series 
in powers of $z$ and integrating one arrives at the result
\bea
&&
\varphi_{pp}^{(2)}(x)=C-\frac{1}{1+4\alpha}\frac{x^2}{2}
+\frac{x^3}{6}
\label{8}\\
&&\phantom{aa}
-\frac{1+16\alpha+96\alpha^2+264\alpha^3+288\alpha^4}{12(1+4\alpha)^3}
\frac{x^4}{4}+O(x^6),
\nonumber
\eea
\noindent
where $C=const$.

The even powers of $x$ evidently do not contribute to the second term
of (\ref{4}). As a result there is only one temperature correction 
originating 
from $Q_2$ which is caused by the term $x^3/6$ and which does not depend on
$\tilde\omega_p$. Substituting (\ref{8}) into the second term of (\ref{4})
one obtains
\beq
\Delta_{T}F_{pp}^{(2)}(a)=F_{pp}^{(0)}(a)\frac{1}{6}
\left(\frac{T}{T_{\mbox{\scriptsize{eff}}}}\right)^4.
\label{9}
\eeq
\noindent
Here $F_{pp}^{(0)}(a)=-\pi^2\hbar c/(240 a^4)$. Note that we neglect the
corrections $O\left[(T/T_{\mbox{\scriptsize{eff}}})^5\right]$.

Consider now $i=1$ in (\ref{7}). In this case both the first and the
second terms in the right-hand side depend on $x$. The second term,
however, is an even function of $x$ and for that reason it does not
contribute to (\ref{4}). Let us expand the quantity $z^2Q_1^{-1}$
in powers of small parameters $\alpha$ and $z$. Integrating the obtained
series between the limits $z=x$ and $z=x_0<1$ we obtain
\beq
\varphi_{pp}^{(1)}(x)=\frac{x^3}{6}+4x^2\alpha\ln x+
\tilde\varphi_{pp}^{(1)}(x),
\label{10}
\eeq
\noindent
where the quantity $\tilde\varphi_{pp}^{(1)}(x)$ contains terms which
do not contribute
to (\ref{4}) or lead to contributions of order 
$(T/T_{\mbox{\scriptsize{eff}}})^5$
or higher. Substituting (\ref{10}) into the second term of (\ref{4})
we get
\beq
\Delta_{T}F_{pp}^{(1)}(a)=F_{pp}^{(0)}(a)\!
\left[\frac{1}{6}\left(\frac{T}{T_{\mbox{\scriptsize{eff}}}}
\right)^4\!
+\frac{30\zeta(3)}{\pi^3}\frac{\delta_0}{a}
\left(\frac{T}{T_{\mbox{\scriptsize{eff}}}}\right)^3\right],
\label{11}
\eeq
\noindent
where $\zeta(3)\approx 1.202$ is the Riemann zeta function.

Now, let us take together
(\ref{9}), (\ref{11})
and the zero temperature contribution           
given by the first term
of (\ref{4}). In \cite{25} the last one   
was calculated up to the fourth order.
Here we add two more orders. The final result is
\bea
&&
F_{pp}(a)=F_{pp}^{(0)}(a)\left\{
\vphantom{\left(\frac{163\pi^2}{73500}\right)\frac{\delta_0^4}{a^4}}
1+\frac{1}{3}\left(\frac{T}{T_{\mbox{\scriptsize{eff}}}}
\right)^4\right.
\label{12}\\
&&
-\frac{16}{3}\frac{\delta_0}{a}
\left[1-\frac{45\zeta(3)}{8\pi^3}\left(
\frac{T}{T_{\mbox{\scriptsize{eff}}}}\right)^3\right]
+\left.\sum\limits_{i=2}^{6}c_i\frac{\delta_0^i}{a^i}
\vphantom{\left(\frac{163\pi^2}{73500}\right)\frac{\delta_0^4}{a^4}}
\right\},
\nonumber
\eea
\noindent
where $c_2=24$, and the other coefficients are
\bea
&&
c_3=-\frac{640}{7}\left(1-\frac{\pi^2}{210}\right),\ 
c_4=\frac{2800}{9}\left(1-\frac{163\pi^2}{7350}\right),
\nonumber\\
&&
c_5=-\frac{10752}{11}
\left(1-\frac{305\pi^2}{5292}+\frac{379\pi^4}{1693440}\right),
\label{12a}\\
&&
c_6=\frac{37632}{13}
\left(1-\frac{1135\pi^2}{9720}+\frac{2879\pi^4}{1358280}\right).
\nonumber
\eea

For $\delta_0=0$ (perfect conductor) Eq.~(\ref{12}) turns into the well
known result [13--15]. It is significant that the first correction of 
mixing finite conductivity and finite temperature is of order 
$(T/T_{\mbox{\scriptsize{eff}}})^3$,
and there are no temperature corrections up to 
$(T/T_{\mbox{\scriptsize{eff}}})^4$ in the
higher conductivity corrections from the second up to the six order.

Analogous calculations can be performed for the configuration
of a sphere (lens) of radius $R$ above a plate starting from the force
\beq
F_{pl}(a)=\frac{k_B TR}{c^2}
\sum\limits_{n=0}^{\infty}{\vphantom{\sum}}^{\prime}
\xi_n^2\!
\int\limits_{1}^{\infty}\!\!
pdp
\ln\frac{Q_1Q_2}{(Q_1+1)(Q_2+1)}.
\label{13}
\eeq
\noindent
This formula is obtained from (\ref{1}) using the proximity force
theorem \cite{26}, $Q_{1,2}$ are defined in (\ref{2}). After 
straightforward calculations, 
using \cite{25} for the zero temperature contribution,
the result is
\bea
&&
F_{pl}(a)=F_{pl}^{(0)}(a)\left\{
\vphantom{\left(\frac{163\pi^2}{73500}\right)\frac{\delta_0^4}{a^4}}
1+\frac{45\zeta(3)}{\pi^3}\left(
\frac{T}{T_{\mbox{\scriptsize{eff}}}}\right)^3
-\left(\frac{T}{T_{\mbox{\scriptsize{eff}}}}\right)^4\right.
\label{14}\\
&&
-4\frac{\delta_0}{a}
\left[1-\frac{45\zeta(3)}{2\pi^3}\left(
\frac{T}{T_{\mbox{\scriptsize{eff}}}}\right)^3
+\left(\frac{T}{T_{\mbox{\scriptsize{eff}}}}\right)^4\right]
+\left.\sum\limits_{i=2}^{6}\tilde{c}_i\frac{\delta_0^i}{a^i}
\vphantom{\left(\frac{163\pi^2}{73500}\right)\frac{\delta_0^4}{a^4}}
\right\},
\nonumber
\eea
\noindent
where $F_{pl}^{(0)}(a)=-\pi^3\hbar cR/(360a^3)$, $\tilde{c}_i=3c_i/(3+i)$,
$c_i$ are defined in (\ref{12a}).
For the perfect conductor $\delta_0\to 0$ 
the known asymptotic behavior \cite{10} is reproduced.

Now we consider space separations $a$ for which 
$T\sim T_{\mbox{\scriptsize{eff}}}$ or even
larger. In this case perturbation theory in 
$T/T_{\mbox{\scriptsize{eff}}}$ does not work.
Let us compute the values of temperature force (\ref{13}) 
numerically in dependence on $a$ for $Al$ surfaces
used in experiments \cite{11,13} with $\omega_p=1.92\times 10^{16}\,$rad/s
\cite{27}, $T=300\,$K, and
$R=100\,\mu$m.  
The numerical results are shown in Fig.~1 by the solid curve. In the
same figure the asymptotic behavior (\ref{14}) is presented by the
pointed line. The dashed line shows the Casimir force at zero
temperature (but with account of finite conductivity). Here, the force was
computed by Eq.~(\ref{13}) in which the sum has been changed into the
integral \cite{19}. 
It is seen that perturbation theory works well within
the range $0.1\,\mu\mbox{m}\leq a\leq 3.5\,\mu$m (note that all six
perturbation orders are essential near the left verge of this interval). 
Starting from
$a=6\,\mu$m the solid line represents the asymptotics at large separations
(temperatures)
\beq
F_{pl}(a)=-\frac{\zeta(3)}{4a^2}Rk_B T\left(1-2\frac{\delta_0}{a}\right).
\label{16}
\eeq
\noindent
This result follows from the term of (\ref{13}) with $n=0$ (the other terms
being exponentially small in 
$T/T_{\mbox{\scriptsize{eff}}}$). For $\delta_0=0$ one obtains
from (\ref{16}) the known expression for perfect conductors [7,13--15].
Finite conductivity corrections of higher orders do not contribute at
large separations.

By way of example, consider the contribution of temperature correction
for the $Al$ sphere and the plate of experiments \cite{11,13} 
at smallest separations ($a\sim 0.1\,\mu$m) as
calculated in \cite{21}. According to the numerical results obtained
by the plasma model
$\Delta_T F_{pl}=F_{pl}(T)-F_{pl}(T=0)\approx 0.03 \,$pN, where $F_{pl}(T)$ is
computed by Eq.~(\ref{13}).
Almost the same result is obtained by the perturbative result of
Eq.~(\ref{14}). These values fall far short of the computational result
$|\Delta_T\tilde{F}_{pl}|\approx 4\,$pN presented in \cite{21}. Although the
computational procedure is not described explicitly in \cite{21} we have been
able to reproduce the value obtained there as follows. 

In \cite{21}, instead of the plasma model, the Drude model was used
at small frequences 
for which the dielectric permittivity on the imaginary axis is 
\beq
\varepsilon(i\xi)=1+\frac{\omega_p^2}{\xi(\xi+\gamma)}.
\label{15}
\eeq
\noindent
Here, $\gamma$ is the relaxation frequency. Substituting the dielectric
permittivity (\ref{15}) into (\ref{13}) and performing computations 
the incorrect values
$|\Delta_T\tilde{F}_{pl}|\approx (2.5-8.5)\,$pN are obtained
when separations decrease from 0.1\,$\mu$m till 0.09\,$\mu$m
(compare with $\approx 4\,$pN in \cite{21}). 
It arises for the following reasons. 

When performing computations, proper allowance must be
made for a critical issue resolved by J.~Schwinger, L.~L.~DeRaad, Jr., and
K.~A.~Milton \cite{19}. In line with \cite{19} the prescription should be
adopted that we take the limit $\varepsilon\to\infty$ before setting
$\xi=0$ in order that the Casimir force between perfect conductors be
obtained from Eqs.~(\ref{1}), (\ref{13}). Otherwise the $n=0$ terms
of (\ref{1}), (\ref{13}), containing $Q_2$, would not contribute, which
would imply incorrect limits both at low and high temperatures \cite{19}.
For a real metal the prescription of \cite{19} is satisfied automatically when
the plasma model (\ref{5}) is used. In the case of Drude model (\ref{15}),
however, the contribution of $Q_2$ in the terms of (\ref{1}), 
(\ref{13}) with $n=0$ is identical zero. 
Because of this, it is impossible to follow
the prescription of \cite{19}. 
 It is necessary to stress that if the prescription
of \cite{19} is not carried out for the calculation of the zeroth term of
(\ref{13}) one would obtain $-\zeta(3)Rk_B T/(8a^2)$ instead of (\ref{16})
(remind that the term containing $Q_2$ does not contribute in this case).
The last expression is evidently incorrect. It is two times smaller than 
the main contribution to (\ref{16}), which is valid for perfect conductor,
and is independent of the actual value of conductivity of the real metal
under consideration. Now it has been evident that the extra contribution of
$\approx$4\,pN discussed above originates from the missing 
contribution of $Q_2$,
when the Drude model is used, and is equal to it by the modulus.
Because of this, Drude model can not be used to provide a correct
extraction of the case of metals from the Lifshits formula for
dielectrics.

In conclusion it may be said that the joint effect of non-zero
temperature and finite conductivity on the Casimir force was examined.
It turned out to be in agreement with the previous knowledge for the
real metals at zero temperature from one side 
and for the perfect conductors at
non-zero temperature from the other. (Note that some of the above results 
related to the  plasma model only were obtained independently in the
recent preprint \cite{28}.) 
The obtained results are the topical ones for the
interpretation of precision measurements of the Casimir force.

\section*{ACKNOWLEDGMENTS}  

The authors are grateful to S.~Reynaud for attracting their attention to
Ref.~\cite{28} and discussion.
G.L.K.\ and V.M.M.\ are indebted 
to Center of Theoretical Sciences and  
Institute of Theoretical Physics of Leipzig University,
where this work was performed, for kind hospitality. G.L.K.
\ was supported by Graduate College on Quantum Field Theory at Leipzig
University.  V.M.M.\ was
supported by Saxonian Ministry for Science and Fine Arts.

\begin{figure}[h]
\caption{
The Casimir force in pN as a function of the surface
separation in configuration of a sphere above a disk. The solid
line represents the computational results obtained by Eqs.~(\ref{5}),
(\ref{13}). The dotted line is calculated by the perturbative
Eq.~(\ref{14}) up to sixth order in relative penetration depth and fourth
order in relative temperature. The dashed line is the zero temperature result.
}
\end{figure}
\twocolumn[\hsize\textwidth\columnwidth\hsize\csname@twocolumnfalse%
\endcsname
{\begin{figure}[h]
\centerline{\epsffile{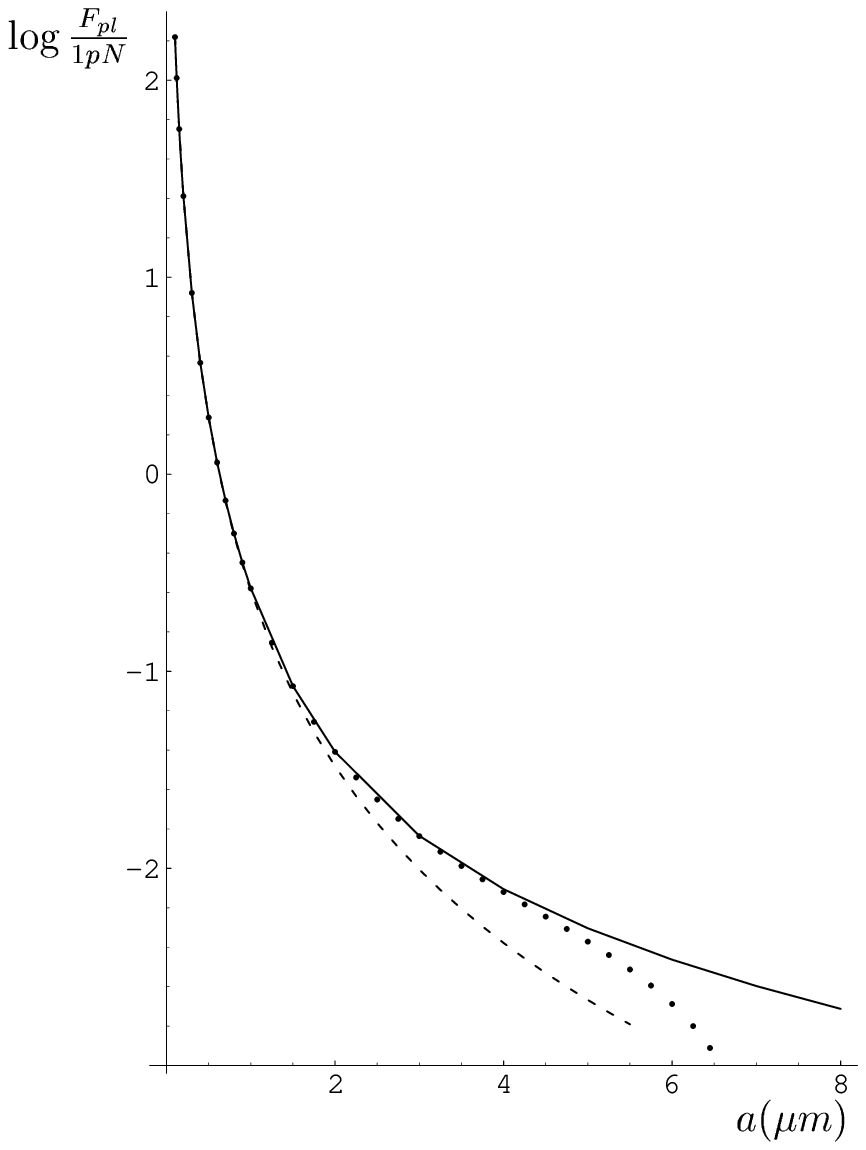} }
\end{figure}}
]

\begin{thebibliography}{99}
\bibitem{1}
M.~Schaden and L.~Spruch,
 Phys. Rev. Lett. {\bf 84}, 459 (2000).
\bibitem{2}
L.~H.~Ford,
 Phys. Rev. A {\bf 58}, 4279 (1998).
\bibitem{3}
R.~Golestanian and M.~Kardar,
 Phys. Rev. Lett. {\bf 78}, 3421 (1997).
\bibitem{4}
M.~Bordag,  K.~Kirsten, and D.~Vassilevich,
 Phys. Rev. D {\bf 59}, 085011 (1999).
\bibitem{5}
I.~Brevik, V.~N.~Marachevsky, and K.~A.~Milton,
 Phys. Rev. Lett. {\bf 82}, 3948 (1999).
\bibitem{6}
I.~Klich,
 Phys. Rev. D {\bf 61}, 025004 (2000).
\bibitem{10}
S.~K.~Lamoreaux, { Phys. Rev. Lett.} 
{\bf 78}, 5 (1997); {\bf 81}, 5475(E) (1998).
\bibitem {11}
U.~Mohideen and A.~Roy, 
{ Phys. Rev. Lett.} 
{\bf 81}, 4549 (1998).
\bibitem {13}
A.~Roy, C.-Y.~Lin, and U.~Mohideen, 
{ Phys. Rev. D} 
{\bf 60}, R111101 (1999).
\bibitem {14}
G.~L.~Klimchitskaya, A.~Roy, U.~Mohideen, and V.~M. Mos\-te\-panenko, 
{ Phys. Rev. A} 
{\bf 60}, 3487 (1999).
\bibitem{15}
E.~M.~Lifshitz,
Sov. Phys. JETP (USA) {\bf 2}, 73 (1956).
\bibitem {16}
E.~M.~Lifshitz and L.~P.~Pitaevskii, 
{\it Statistical Physics, Part 2} (Pergamon Press,
Oxford, 1980).
\bibitem {17}
J.~Mehra, Physica {\bf 37}, 145 (1967).
\bibitem {18}
L.~S.~Brown and G.~J.~Maclay, Phys. Rev. {\bf 184}, 1272 (1969).
\bibitem {19}
J.~Schwinger, L.~L.~DeRaad,~Jr., and K.~A.~Milton,
{ Ann. Phys.} (N.Y.) {\bf 115}, 1 (1978).
\bibitem{20}
M.~Bordag, B.~Geyer, G.~L.~Klimchitskaya,  
and V.~M. Mos\-te\-panenko,
{ Phys. Rev.} D {\bf 58}, 075003  (1998).
\bibitem{21}
V.~B.~Svetovoy and M.~V.~Lokhanin,
e-print quant-ph/0001010.
\bibitem {22} 
V.~M.~Mostepanenko and N.~N.~Trunov, {\it The Casimir Effect 
and Its Applications}
(Clarendon Press, Oxford, 1997).
\bibitem{23}
A.~Lambrecht and S.~Reynaud,
Eur. Phys. J. D {\bf 8}, 309 (2000).
\bibitem{24}
S.~K.~Lamoreaux, { Phys. Rev. A} 
{\bf 59}, R3149 (1999).
\bibitem {25}
V.~B.~Bezerra, G.~L.~Klimchitskaya, and V.~M.~Mostepanenko,
e-print quant-ph/9912090.
\bibitem {26}
J.~Blocki, J.~Randrup, W.~J.~Swiatecki,
 and C.~F.~Tsang,
{ Ann. Phys.} (N.Y.) {\bf 105}, 427 (1977).
 \bibitem {27}
{\it Handbook of Optical Constants of Solids},
edited by E.D.~Palik (Academic Press, New York, 1998).
 \bibitem {28}
C.~Genet, A.~Lambrecht, and S.~Reynaud,
e-print quant-ph/0002061.
\end{thebibliography}
\end{document}